\documentclass{article}
\usepackage{graphicx} 
\usepackage[margin=1in]{geometry} 
\usepackage{helvet} 
\usepackage{authblk} 

\PassOptionsToPackage{hyphens}{url}\usepackage{hyperref}
\title{The Closure Challenge: a benchmark task for machine learning in turbulence modelling}

\author[1*$\dagger$]{Ryley McConkey}
\author[2*]{Tyler Buchanan}
\author[1]{Tess Smidt}
\author[1]{Abigail Bodner}
\author[2]{Richard Dwight}
\author[3]{Paola Cinnella}
\affil[1]{Center for Computational Science and Engineering, Massachusetts Institute of Technology, United States}
\affil[2]{Aerodynamics Group, Faculty of Aerospace Engineering, TU Delft, The Netherlands}
\affil[3]{
Institut Jean Le Rond D’Alembert,
Sorbonne University, France}
\affil[*]{Equal contribution.}
\affil[$\dagger$]{Corresponding author: \texttt{rmcconke@mit.edu}.}

\usepackage[framemethod=default]{mdframed}
\date{}
\begin{document}
\maketitle

\noindent\textbf{Abstract}\\

\noindent We introduce a field-wide benchmark challenge for machine learning in Reynolds-averaged Navier-Stokes (RANS) turbulence modelling. Though open-source datasets exist for training data-driven turbulence closure models, the field has been notably lacking a standard benchmark metric and test dataset. The Closure Challenge is a curated collection of open-source datasets and evaluation code that remedies this problem. We provide a variety of high-fidelity training data in a standardized format, including mean velocity gradients. The test cases (periodic hills, square duct, and NASA wall-mounted hump) evaluate Reynolds number and geometry generalization, two key issues in the field. We present results from three early submissions to the challenge. This is an ongoing challenge, intended to continuously spur innovation in machine learning for turbulence modelling. Our goal is for this benchmark to become the standard evaluation for new machine learning frameworks in RANS. The Closure Challenge is available at: \url{https://github.com/rmcconke/closure-challenge-benchmark}.

\section{Introduction}
The field of ML augmented RANS modelling has seen significant interest for at least a decade. Many methodologies have been proposed. However, a critical problem slowing progress in the field is the absence of an open-source benchmark dataset with clear evaluation criteria. In order to compare a new technique against an existing technique, significant effort is required. We aim to eliminate this required effort and greatly accelerate progress in the field by implementing a benchmark dataset for ML in RANS. This mirrors the use of field-wide benchmarks in many other scientific machine learning domains,
such as protein structure prediction \cite{casp}, materials property prediction \cite{matbench},
catalyst discovery \cite{oc20}, PDE surrogate modelling \cite{pdebench}, and weather forecasting
\cite{weatherbench}.

Our goal is to create a challenging dataset that represents the actual state of ML-augmented RANS turbulence modelling. We aim to propose challenging generalization tasks, with the goal that over time, techniques which generalize better will rise to the top of the leaderboard. We do not want to cast the field in an overly optimistic light; we want to provide a hard challenge that will motivate new ideas in the field.

This is an ongoing challenge. It is not associated with any particular conference or event. This running leaderboard aims to summarize the state of the art in the field of ML for RANS turbulence modelling. We aim to make it a standard requirement to test future methodologies on this dataset, thereby greatly accelerating progress in this field.  

The primary source of information for this challenge is the github repository: \url{https://github.com/rmcconke/closure-challenge-benchmark}. This document outlines the task and philosophy of the challenge, as well as the current results. We intend to update it periodically as more submissions become available.

\section{Challenge task and rules}
The benchmark task is to predict the flow field for a series of test cases on a specified CFD mesh. All other decisions are left to the submitter.

Notably, we leave the training and validation dataset up to the submitter. We would prefer having a standardized training and validation dataset in order to separate the performance of a modelling technique from the data it was trained on. However, the complexities associated with implementing current techniques, and previous lack of a field-wide benchmark mean that each research group has their own code, training datasets, and local infrastructure. Therefore, to lower the barrier to entry, we only standardize the test dataset. You are free to use your own training and validation data.

\subsection{Test cases}
The test cases cannot be used in any way at training time. You can train on similar flows to the test cases (for example, different parametric variations of the periodic hills case). More details of the provided datasets are given in Section~\ref{sec:datasets}.

\begin{mdframed}[linewidth=1pt, linecolor=black, backgroundcolor=white, innertopmargin=10pt, innerbottommargin=10pt]
The only strict rule with this challenge is: \textbf{it is forbidden to train or validate on any of the test cases.} 

\noindent As of March 2026, the test cases are:
\begin{itemize}
    \item \textbf{Periodic hills:} ($Re=5600$)
    \begin{itemize}
        \item $\alpha = 1.5$, $L=13.929H$, $h=4.048H$
        \item $\alpha = 1.5$, $L=13.929H$, $h=2.024H$
        \item $\alpha = 1.5$, $L=4.071H$, $h=4.048H$
        \item $\alpha = 1.5$, $L=4.071H$, $h=2.024H$
    \end{itemize}
    \item \textbf{Square duct:}
    \begin{itemize}
        \item $Re_\tau =360$, $AR=1$
        \item $Re_\tau=360$, $AR=3$
        \item $Re_\tau =180$, $AR=14$
    \end{itemize}
    \item \textbf{NASA wall-mounted hump.}
\end{itemize}
Please check the challenge website for a current list of test cases.

\end{mdframed}

\subsection{Submission}
You must submit your predictions on the provided meshes for each of the test cases. The Python evaluation package for the challenge is open-source and available on pypi. Instructions for using the evaluation package to preview your score are given at \url{https://github.com/rmcconke/closure-challenge}.

Please check the main challenge github page for current instructions on how to submit your predictions. As of March 2026, submissions are via email. 

\subsection{Scoring}
As of March 2026, the score is calculated as follows:
\begin{equation}
    \mathrm{Score} = \frac{1}{N_{\mathrm{cases}}} \sum_{c} \frac{1}{N_c \, \overline{|\mathbf{U}|}_c} \sum_{i=1}^{N_c} \left| \tilde{\mathbf{U}}_i - \mathbf{U}_{{\mathrm{true}},i} \right|
\end{equation}
where $c$ indexes the test cases, $N_{\mathrm{cases}}$ is the total number of test cases, $N_c$ is the number of evaluation points in test case $c$, $|\cdot|$ denotes the Euclidean vector magnitude, $\tilde{\mathbf{U}}_i$ is the predicted velocity vector at point $i$, $\mathbf{U}_{{\mathrm{true}},i}$ is the reference (DNS/LES) velocity vector, and $\overline{|\mathbf{U}|}_c = \frac{1}{N_c}\sum_{i=1}^{N_c} |\mathbf{U}_{{\mathrm{true}},i}|$ is the mean velocity magnitude over test case $c$. The scaling by $\overline{|\mathbf{U}|}_c$ ensures that each test case contributes on a comparable scale regardless of its characteristic velocity. The overall score can be interpreted as the average fractional velocity error across all test cases; for example, a score of 0.05 indicates that predictions are off by approximately 5\% of the mean velocity magnitude on average. A lower score indicates better agreement with the reference data.

\section{Datasets}\label{sec:datasets}

The following fields are available for each of the datasets:

\begin{itemize}
    \item RANS predictions with the $k$-$\omega$ SST turbulence model
    \item DNS or LES "ground truth" data, including mean velocity gradients.
\end{itemize}

We include the velocity gradients, as we note that many current frameworks require this information at training time. Since there is no standardized training dataset for this challenge, we provide this data purely for convenience and to lower the barrier to entry for the challenge. 

The full list of included datasets is continuously updated on the challenge github page. As of March 2026, the included datasets are:
\begin{figure}
    \centering
    \includegraphics[width=0.7\linewidth]{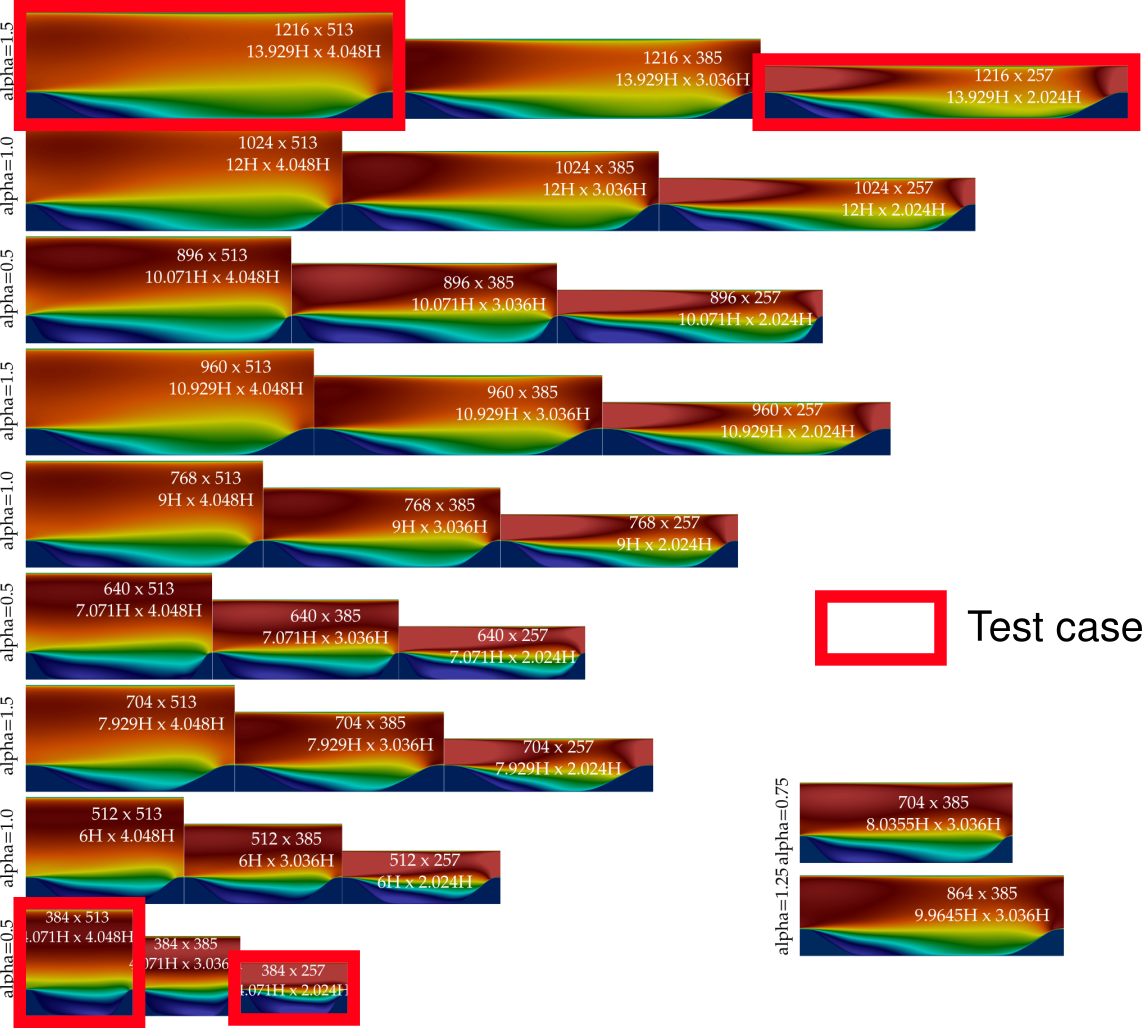}
    \caption{Cases for the periodic hills dataset (reproduced from~\cite{xiao} with red boxes added).}
    \label{fig:periodic_hills}
\end{figure}
\begin{itemize}
    \item \textbf{2D flows}
    \begin{itemize}
    \item Periodic hills, 29 parametric variations ($Re=5600$) \cite{xiao}
    \item Periodic hills, $Re=10595$ \cite{tmr}
    \item Square and rectangular ducts \cite{Vinuesa_duct1,Vinuesa_duct2,Vinuesa_duct3,Vinuesa_duct4}
    \item Curved backward-facing step \cite{tmr}
    \item NASA wall-mounted hump \cite{tmr}
\end{itemize}
\item \textbf{3D flows}
\begin{itemize}
    \item Wing-body junction flow, $Re=1.15 \times 10^5$ \cite{wingbody1,wingbody2}
    \item Ahmed body wake, $Re=7.60 \times 10^5$ \cite{ahmed1,ahmed2,ahmed3}
    \item FAITH hill flow, $Re=5.00 \times 10^5$ \cite{tmr}
\end{itemize}
\end{itemize}
Figure~\ref{fig:periodic_hills} shows the test cases chosen for the periodic hills dataset.

\section{Current status and future directions}
The initial challenge involved submission from three groups. This submission and initial evaluation was synchronized with the 3rd European Research Commission on Flow, Turbulence, and Combustion (ERCOFTAC) Machine Learning for Fluid Dynamics workshop. However, this is a continuously running benchmark, not associated with any event. The github page will be updated with any new submissions. Table~\ref{tbl:leaderboard} presents the leaderboard as of March 2026. Details from each of the early submitters are available on the github page.

\begin{table}[ht]
\centering
\renewcommand{\arraystretch}{1.2}
\caption{Leaderboard as of March 2026.}\vspace{1mm}
\label{tbl:leaderboard}
\begin{tabular}{clc}
\hline
Rank & Authors & Overall \\
\hline
1 & Reissmann, Fang, and Sandberg \cite{reissmann,WEATHERITT201622} & 0.0595 \\
2 & Wu and Zhang \cite{wu}                         & 0.0624 \\
3 & Montoya, Oulghelou, and Cinnella \cite{montoya}$^*$ & 0.0779 \\
\hline
\multicolumn{3}{l}{\small $^*$Pretrained model from \cite{montoya} only, without fine-tuning on the challenge datasets.}
\end{tabular}
\end{table}

As of March 2026, the challenge is implemented and running for the 2D flows. 3D data is currently available on the challenge page. We intend to add an additional challenge and leaderboard for 3D flows in the future.

\bibliography{references}
\end{document}